\documentstyle[aps,prd]{revtex}

\begin{document}

\draft
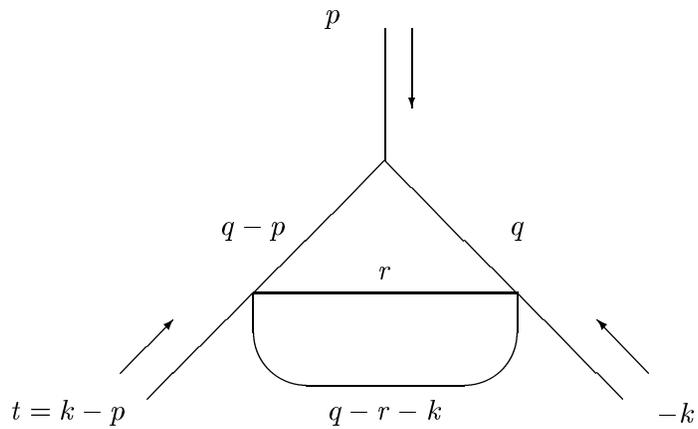
\title{Solutions for a massless off-shell two-loop three-point vertex} 

\author{Alfredo T. Suzuki and Alexandre G. M. Schmidt}
\address{Instituto de F\'{\i}sica Te\'orica - R.Pamplona, 145, S\~ao Paulo, SP
CEP 01405-900, Brazil }
\date{\today}

\maketitle

\begin{abstract} 
Negative dimensional integration method (NDIM) seems to be a very 
promising technique for evaluating massless and/or massive Feynman 
diagrams. It is unique in the sense that the method simultaneously 
gives solutions in different regions of external momenta. Moreover, 
it is a technique whereby the difficulties associated with performing 
parametric integrals --- the standard approach --- are transferred to a
simpler solving of a system of linear algebraic equations. Employing
this method, we calculate a massless two-loop three point vertex with
all the external legs off-shell. Then NDIM approach allows us to obtain
twenty-one distinct new power series representations for the integral
in question. In order to verify the correctness of our results, we
consider five particular cases where either two of the external legs are put
on-shell, or one of them amputaded or one exponent of the propagators is set to
zero,  and compare our results thus obtained with the ones calculated
with standard methods in positive dimension.
\end{abstract}

\def\be{\begin{equation}}
\def\ee{\end{equation}}
\def\beq{\begin{eqnarray}}
\def\eeq{\end{eqnarray}}
\def\s{\sigma}
\def\G{\Gamma}
\def\F{_2F_1}
\def\an{analytic}
\def\ac{\an{} continuation}
\def\hsr{hypergeometric series representations}
\def\hf{hypergeometric function}
\def\ndim{NDIM}
\def\quarto{\frac{1}{4}}

 \def\half{\frac{1}{2}}

\section{Introduction.}

Everyone who has ever studied perturbative quantum field theory (QFT) has
come face to face with Feynman integrals and knows all too well that
evaluating these loop integrals is often a hard task. Books on
QFT\cite{zuber} generally teach us how to deal with them either in 
terms of the so-called $\alpha$-parametrization or by using Feynman
parameters for the propagators. This approach, contrary to what we may
think, is perhaps one of the most laborious techniques to solve them
because the number of parametric integrals increases with the number of
propagators. Moreover, the added parametric integrals are often very
difficult to solve exactly even if there are no massive particles in
the intermediate states. There are several techniques that have been
developed over the years in order to solve Feynman integrals, and we
can just mention a few ones: the Mellin-Barnes' representation of
massive propagators\cite{boos}, the Gegenbauer polynomial approach in
configuration space and integration by parts\cite{russo} and some
others\cite{kreimer}. Yet in this arena, the simpler the method the
better. A novel approach that was suggested some years ago has to do
with the use of negative dimensions and seems to be a promising
technique.   

In mathematical physics, one of the most powerful and useful principles
is that of \ac{}. It is the underlying principle that allowed, for
instance, 't Hooft {\it et al} \cite{thooft} to develop the elegant
technique of dimensional regularization (DREG). Besides a great number
of other uses in Physics, there is one that particularly is of interest
and that most concerns us here: It is that of negative dimensional
integration method (NDIM)\cite{halliday,halliday2} which combines the
two amazing features in its heart: The principle of \ac{} and the
technique of DREG. What is the advantage of such a combination over the
plain DREG? The latter one allows us to calculate Feynman diagrams in
the analytically continued $D$-dimensional space, while the former
allows us to greatly simplify the technical difficulties associated
with performing parametric integrals, just by analytically extending
$D$ over negative values. This simplification comes about in virtue of
the polynomial nature of the integrands. In a few words, this character
emerges from the equivalence between negative dimensional bosonic
integration and positive dimensional fermionic
integration\cite{halliday2}.   

What is the price one pays for working them out in negative dimensions?
Basically, the technical difficulties that arise are solving many systems of
linear algebraic equations, performing gaussian/gaussian-type integrals and
dealing with multi-indexed power series. For the former two, one can just ask
whether could it be simpler than these? Yet the difficult part is not absent:
the laborious piece comes in the form of multi-indexed power series. However,
looking at it from the encouraging side, we can say that it allows us a
standardized representation for Feynman integrals in terms of power series.  

In a previous paper\cite{lab} we carried out what we, as
theoreticians, call a ``lab-test'', that is, with a new approach one
studies a well-known system. Employing \ndim{} we calculated a massless
two-loop three point vertex, keeping two of the external legs on-shell.
The \ndim{} technique led us to discover twelve different ways in which
the result could be written down, i.e., a twelve-fold degeneracy for
that particular integral. We also considered some two-loop self-energy
diagrams for a massless theory\cite{flying}.

Of course \ndim{} is not the only technique to calculate Feynman integrals.
Recently, Fleischer {\it et al} studied asymptotic expansions of some
two-loop vertex\cite{fleischer} and Frink {\it et al} gave results for a
general massive two-loop three point vertex\cite{kreimer}. Ussyukina and
Davydychev\cite{letb} calculated the Feynman diagram we will study in this
paper but they did give the result in terms of only two dimensionless
variables. On the other hand, we will write down twenty-one results in
terms of not only two combinations of external momenta, but rather in many
different ratio combinations for the external momenta . Our aim here is not
to make any numerical calculations, of course, but analytical ones. The 
interested reader in the two-loop calculation ``technology'' in QFT can
consult a good review on this subject by Davydychev\cite{davyd}. 

The outline for our paper is as follows: in section 2 we calculate the
vertex in question in euclidean space, in section 3 we consider five 
special cases where only one external leg is off-shell and in the last
one, section 4, we conclude the work. 

\section{Off-Shell Two-Loop Vertex.}

This computation is performed following the few simple steps outlined in
\cite{lab,flying}. First of all, let us consider the integral,

\be I = \int\int d^D\!r\; d^D\!q \;\;\exp\left[-\alpha q^2 -\beta
(q-p)^2 -\gamma r^2 -\omega (q-r-k)^2\right] ,\ee
which corresponds to the diagram of figure 1.


The general solution for the integral in negative $D$, defined by

\be \label{Indim} J_{NDIM} = \int\int d^D\!q\;d^D\!r
\;(q^2)^i\left[(q-p)^2\right]^j (r^2)^l\left[(r-q+k)^2\right]^m , \ee
is given by the multiple series,
\be \label{geral} S_{NDIM} = {\cal G}(i,j,l,m;D) \sum_{n_1,...,n_9=0}^\infty
\frac{(p^2)^{n_1+n_2} (k^2)^{n_3}(t^2)^{n_4}} {n_1!n_2!n_3!n_4!n_5!n_6!
n_7!n_8!n_9!} ,\ee
where
\[ {\cal G}(i,j,l,m;D) = (-\pi)^D\G(1+i)\G(1+j)\G(1+l)\G(1+m)\G(1-\s-\half D),
\] 
and for convenience we use the definition $\s=i+j+l+m+D$. The system one must
solve is, 
\be \label{sys2}\left \{ \begin{array}{rcl}
n_1+n_2+n_3+n_5+n_6&=&i \\[.25cm]
n_1+n_2+n_4+n_7+n_8&=&j \\[.25cm]
n_1+n_3+n_4+n_5+n_7+n_9&=&l \\[.25cm]
n_2+n_3+n_4+n_6+n_8+n_9&=&m \\[.25cm]
n_1+n_2+n_3+n_4&=&\s . \end{array} \right. 
\ee

It is an easy matter to see that this system is composed of five equations
with nine ``unknowns'' (the sum indices), so that it cannot be solved unless
it is done in terms of four arbitrary ``unknowns''. These, of course will
label the four remnant summations, which means that the answer will be in
terms of a fourfold summation series. From the combinatorics, it is a
straightforward matter to see that there are many different ways we can
choose those four indices; indeed, we can choose $C_5^9 = 126$
different ways. In other words, what we need to do is to solve $126$
different systems. Of these, $45$ are unsolvable systems, i.e., they are
systems whose set solution is empty. There remains therefore $81$ which has
non-trivial solutions. Of course, the trivial solutions are of no interest at
all. However, the non-trivial solutions generate a space of functions with
different basis, characterized by their functional variable, according to the
different possibilities allowed for ratios of external momenta. Each basis is
a solution for the pertinent Feynman integral, which is connected by \ac{} to
all other basis defined by the other sets of solutions. We remind
ourselves that a basis that generates a given space can be composed of one or
several linearly independent functions combined in what is called linear
combination. 

Each representation of the Feynman integral will be given by a basis of
functions generated by the solutions of the systems\cite{suzuki2}. Of course,
only linearly independent and non-degenerate solutions are relevant to define a
basis.

It can be easily seen that the diagram we are dealing with here is symmetric
under the exchange of external momenta $k^2 \leftrightarrow t^2$. This
symmetry is reflected by the systems we have to solve, and the solutions
display this fact. Therefore, solutions within this category will be given
only once, that is, the ones which can be obtained by symmetry will
not be written down explicitly.

\subsection{Two Variables.}

With the solutions in hands --- it is an easy matter to write down a computer
program to solve all the systems --- and the general form of the results
(\ref{geral}), we can start to build the power series representations of the
Feynman graph. 

We begin our analysis of the solutions for the systems by looking at the
simpler ones having two variables, defined by ratios of external momenta.  Of
course, there are in fact four sums but two of them have unity argument,
making it possible for us to actually sum the pertinent series as we shall
shortly see. The variables are,  

\be
\begin{array}{l}
(x,y) ,\;\; (z,y^{-1}),\\[.25cm]
(x^{-1},z^{-1}),
\end{array}
\ee
where
we define the dimensionless ratios
\beq
x&=&\frac{p^2}{k^2},\nonumber\\[.25cm]
y&=&\frac{t^2}{k^2},\\[.25cm]
z&=&\frac{p^2}{t^2}\nonumber .
\eeq

Note that the second pair of variables above is exactly symmetric to the
first one by the interchange $k^2 \leftrightarrow t^2$, which means that we
will not write this second solution explicitly. Also, each of the three
pairs above appears twelve times among the total of 81 systems with
non-trivial solutions. Each of these is therefore twelve-fold degenerate
just like in the on-shell case calculated in \cite{lab}. A way of
expressing the first solution in positive $D$ is given by 
\beq \label{2vars} S_1^{AC}
&=& \pi^D P_1^{AC} (p^2)^{i+j+\half D} (k^2)^{l+m+\half D}\!\!\!\!\!\!
\sum_{n_4,n_6,n_7,n_9=0}^\infty
\frac{(x)^{n_9}(y)^{n_4}}{n_4!n_6!n_7!n_9!}
\nonumber\\[.25cm] 
&\times &\frac{(-l-m-\half
D|n_4+n_9)(\half D+i|n_4+n_7+n_9)}{(1+j+l+\half D|-n_4+n_6-
n_7) }\nonumber\\[.25cm] 
&\times &\frac{(-1)^{n_7}}{(1+i-l|n_4-n_6+n_7+n_9)(1-j-l-m-D|n_4-n_6)},\eeq  
where
\beq P_1^{AC} &=&  (-i|-j-\half D)(-j|j+l+\half D)(-l|l+m+\half D) \nonumber\\[.25cm]
& \times &(-m|-l-\half D)(l+m+D|j)(\s +\half D|i-\s) .\eeq

Using the property $(a|b+c) = (a+b|c)(a|b)$ and the well-known
summation formula\cite{bateman,rainville} of Gauss' \hf{} $\F$ with unity
argument,
\be \label{2f1} \F(a,b;c|1) = \frac{\G(c)\G(c-a-b)}{\G(c-a)\G(c-b)}
,\ee 
we can sum the series in $n_6$ and in $n_7$ above. We then get,

\beq \label{S1} S_1^{AC} &=& \pi^D(p^2)^{i+j+\half D} (k^2)^{l+m+\half
D}P_1^{AC} \!\!\sum_{n_4,n_9=0}^\infty
\frac{(x)^{n_9}(y)^{n_4}}{n_4!n_9!} \nonumber\\[.25cm] 
& \times &\frac{(-l-m-\half D|n_4+n_9)(\half D+i|n_4+n_9)} 
{(1-j-l-m-D|n_4)(1+i+j+\half D|n_9)} \nonumber\\[.25cm]
& \equiv & \pi^D(p^2)^{i+j+\half D} (k^2)^{l+m+\half
D}P_1^{AC}\nonumber \\[.25cm]
& \times & F_4\left(\left. \begin{array}{rl}
-l-m-\half D,&\half D+i\\[.25cm]
1+i+j+\half D,&1-j-l-m-D\end{array}\right|x,\;y\right).
\eeq 

The remnant double summation in $n_4$ and $n_9$ is by definition the
Appel's $F_4$ \hf{}\cite{bateman,appel} of two variables. In the
particular case where $i=j=l=m=-1$ we can simplify even more this
result by using a reduction formula\cite{bateman} which relates
the $F_4$ function to the gaussian \hf{} $\F$,  

\beq \lefteqn{F_4 \left( \alpha,\beta;1+\alpha-\beta,\beta
\left|\frac{-u}{(1-u)(1-w)}, \frac{-w}{(1-u)(1-w)}\right)\right.}\nonumber
\\[.25cm]
&&=(1-w)^{\alpha}\;\;\F\left(\alpha,\beta;1+\alpha-\beta\left|\frac{-u(1-w)}{1-u}
\right)\right. .\eeq

For this special case,
\beq S_1^{AC}(-1,\cdots, -1) &=& \pi^D\frac{1}{(k^2)^{2-D/2}}\left(\frac{1}{p^2}-
\frac{1}{k^2}\right)^{2-\half D}\nonumber\\[.25cm]
&\times&\frac{\G(D-3)\G^2(2-\half D)\G^3(\half D-1)}
{\G(D-2)\G(\frac{3}{2}D-4)} \nonumber\\[.25cm] 
& \times &\;\F\left(2-\half D, \half D-1;4-D\left|
\frac{kp}{p^2}\frac{p^2-kp}{k^2-kp}\right. \right) .\eeq  

We note that in order to regularize the divergences we can follow the standard
procedure of dimensional regularization\cite{zuber}.

The next solution (third one) also gives double series. Following
the same steps we can sum two series of unity argument and the remaining two are by
definition the Appel's \hf{} $F_4$,

\beq \label{S2} S_2^{AC} &=& \pi^D(p^2)^\s P_2^{AC}\nonumber\\[.25cm]
& \times & F_4\left(-\s, -l-m-\half D; 1+j-\s,
1+i-\s\left|x^{-1},z^{-1}\right.\right),\eeq 
where
\beq P_2^{AC} &=& (-i|\s)(-j|\s)(-l|-m-\half D)(-m|l+m+\half D) \nonumber\\[.25cm]
& \times &(l+m+D|-l-\half D)(\s+\half D|-2\s-\half D) .\eeq

Again, in the particular case where all the exponents of propagators are minus
one, this $F_4$ function reduces to a gaussian \hf{}, too. Using another
reduction formula\cite{bateman}, namely,

\beq
\lefteqn {F_4\left(\alpha, \beta;\;\beta,
\beta\left|\frac{-u}{(1-u)(1-w)},\frac{-w}{(1-u)(1-w)}\right)\right.}
\nonumber\\[.25cm] 
&& = (1-u)^{\alpha}(1-w)^{\alpha}\;\F\left(\alpha,1+\alpha-
\beta;\;\beta \left|uw\right)\right. ,\eeq
we get,
\beq S_2^{AC} &=& \pi^D (p^2)^\s \left[\frac{(p^2-kp)kp}
{k^2t^2}\right]^{2-D/2}\;P_2^{AC}  \nonumber\\[.25cm]
&\times& \;\F\left(2-\half D, \half
D-1;4-D\left|\frac{\left(k^2-kp\right)^2}{k^2t^2}\right)\right. .\eeq  



\subsection{Three Variables.}

In a manner similar to the previous results, there are solutions which have
three remaining variables, meaning that one of the series with unity argument
is summed out. There are six sets of these, determined by their variables, each
appearing four times, i.e., a fourfold degeneracy. Just to keep our accounting
straight, $4 \times 6 = 24$ systems yielding solutions with three variables.
These, added to the $36$ of the previous subsection, gives us $60$ from
the total of $81$ non-trivial systems.

The solutions within this category have functional dependencies given by:

\beq
\begin{array}{rl}
(x,\;x,\;y),& (z,\;z,\;y^{-1}) \nonumber \\[.25cm]
(x,\;y,\;y),& (z,\;y^{-1},\;y^{-1})\nonumber \\[.25cm]
(x^{-1},\;x^{-1},\;z^{-1}),& (z^{-1},\;z^{-1},\;x^{-1}) \nonumber
\end{array}
\eeq

Note that the above triplets are conveniently arranged into pairs connected by
the symmetry $k^2 \leftrightarrow t^2$, so that in the following, only three of
them will be dealt with specifically.
 
We list below the triple power series representations provided by \ndim{}, 

\beq \label{S3} S_3^{AC} &=& f_3
\sum_{n_1,n_2,n_3=0}^\infty  
\frac{(z)^{n_1+n_2}(y^{-1})^{n_3}}{n_1!n_2!n_3!}\frac{(m+\half
D|n_1)}{(1-i-j-\half D|n_1+n_2)}\nonumber \\[.25cm]
& \times &\frac{(l+\half D|n_2)(-i|n_1+n_2+n_3)(-\s|n_1+n_2+n_3)}{
(l+m+D|n_1+n_2)(1-i-l-m-D|n_3)} ,\eeq
where 
\beq f_3 &=& \pi^D(t^2)^{\s}(-j|\s)(-l|l+m+\half D)(-m|l+m+\half D)\nonumber\\[.25cm] 
& \times &(l+m+D|i+j-l-m-\half D)(\s+\half D|-2\s-\half D)\nonumber,
\eeq

\beq \label{S4} S_4^{AC} &=& f_4
\sum_{n_7,n_8,n_9=0}^\infty  
\frac{(y^{-1})^{n_7+n_8}(z)^{n_9}}{n_7!n_8!n_9!} \frac{(m+\half D|n_7)}{(1+i+j+\half
D|n_9)}\nonumber \\[.25cm] 
& \times &\frac{(l+\half D|n_8)(\s+\half D|n_7+n_8+n_9)(i+\half D|n_7+n_8+n_9)}{
(l+m+D|n_7+n_8)(1+i+l+m+D|n_7+n_8)} ,\eeq
where 
\beq f_4 &=& \pi^D(k^2)^{\s}x^jz^{i+\half D}(-i|-j-\half D)(-j|j+l+\half
D)\nonumber\\[.25cm]  
& \times &(-l|-i-m-D)(-m|i+m+\half D)(l+m+D|-l-\half D)\nonumber,\eeq

\beq \label{S5} S_5^{AC} &=& f_5
\sum_{n_4,n_7,n_8=0}^\infty  
\frac{(x^{-1})^{n_7+n_8}(z^{-1})^{n_4}}{n_4!n_7!n_8!} \frac{(l+\half D|n_8)} {
(1-j+\s|n_7+n_8)}\nonumber \\[.25cm]  
& \times & \frac{(m+\half D|n_7)(-j|n_4+n_7+n_8)(i+\half D|n_4+n_7+n_8)}{(\s-i-j|n_7+n_8)(1+i-\s|n_4)},\eeq
where 
\beq f_5 &=& \pi^D(k^2)^{\s}x^j(-i|\s)(-l|l+m+\half D)(-m|l+m+\half D)\nonumber\\[.25cm]  
& \times &(\s+\half D|i-\s)(l+m+D|i+2j-2\s)\nonumber .\eeq

It is not difficult to note that $f_5$ has only one pole in the particular case
when $i=j=l=m=-1$ and $D=4$. {\it What is going on here?} Looking at the series
we observe that there is a
factor $(1+i-\s|n_4)$ in the denominator. For the special case in question it
gives,
\[ \frac{1}{(1+i-\s|n_4)} = \frac{\G(1+i-\s)}{\G(1+i-\s+n_4)} =
\frac{\G(4-D)}{\G(4-D+n_4)} , \] 
so that the only term which is not singular is the first one, $n_4=0$, while
the others become divergent for $D=4$. Then, in fact, we have a double pole as
it should be.  

\subsection{Four Variables.} 

Lastly, we consider in this subsection solutions of systems of linear algebraic
equations leading to four variables, or better, fourfold summation with four
variables. There are $21$ of these solutions, which completes the total of $81$
non trivial solutions for the systems. Again, the functional variables are
given paired with their corresponding symmetries, as follows:

\beq
\begin{array}{rl}
(z^{-1},\;z^{-1},\;x^{-1},\;x^{-1}), & \nonumber \\[.25cm]
(z,\;z,\;y^{-1},\;y^{-1}), & (x,\;x,\;y,\;y) \nonumber\\[.25cm]
(y,\;y,\;z,\;y^{-1}), & (y^{-1},\;y^{-1},\;x,\;y) \nonumber \\[.25cm]
(x,\;y,\;z,\;y^{-1}), & \nonumber\\[.25cm]
(x,\;y,\;x^{-1},\;z^{-1}), & (z,\;y^{-1},\;z^{-1},\;x^{-1}) \nonumber\\[.25cm]
(z,\;z,\;x^{-1},\;z^{-1}), & (x,\;x,\;z^{-1},\;x^{-1}) \nonumber \\[.25cm]
(z,\;y^{-1},\;z^{-1},\;z^{-1}), & (x,\;y,\;x^{-1},\;x^{-1}) \nonumber\\[.25cm]
\end{array}
\eeq

To get the accounting straight, let us again mention that the first three
appear just once and the remaining nine appear twice, totalling the needed $21$
of this category.

The next two solutions appear just one time; the first one being given by

\beq \label{S6} S_6^{AC} &=& f_6 \sum_{\{n_i\}=0}^\infty \frac
{(z^{-1})^{n_5+n_6}(x^{-1})^{n_7+n_8}}{n_5!n_6!n_7!n_8!}\frac{(l+\half
D|n_6+n_8)}{(1-i+\s|n_5+n_6)} \nonumber\\[.25cm]
& \times & \frac {(m+\half D|n_5+n_7)(\s+\half
D|n_5+n_6+n_7+n_8)}{(1-j+\s|n_7+n_8)}\;,
\eeq
where
\beq
f_6 &=& \pi^D \left(\frac
{k^2t^2}{p^2}\right)^{\s}\;z^i\;x^j\;(-i|-l-m-D)\nonumber\\[.25cm]
& \times & (-j|-l-m-D)(-l|l+m+\half D)(-m|l+m+\half D),\nonumber
\eeq
 
and

\beq \label{S7} S_7^{AC} &=& f_7
\sum_{\{n_i\}=0}^\infty  
\frac
{(z)^{n_1+n_2}(y^{-1})^{n_7+n_8}}{n_1!n_2!n_7!n_8!}\frac {(-j|n_1+n_2+n_7+n_8)}
{(1-j+\s|n_7+n_8)} \nonumber \\[.25cm]
& \times & \frac{(l+\half D|n_2+n_8)(m+\half D|n_1+n_7)}
{(1-i-j-\half D|n_1+n_2)} ,
\eeq 
where 
\beq f_7 &=& \pi^D(k^2)^{\s}\;y^j\;(-i|-l-m-D)(-l|l+m+\half D)\nonumber\\[.25cm]
& \times & (-m|l+m+\half D)(\s+\half D|i+j-\s)\nonumber ,\eeq

The next set of five solutions are such that each one of them appears twice but
with exponents of propagators interchanged, i.e., $l \leftrightarrow m$. This
means that the space of functions here is generated by two linearly independent
basis functions, and the series representation for the Feynman integral will be
given by
$$
S^{AC}_{r} = S^{AC(1)}_{r} + S^{AC(2)}_{r}, \;\mbox {r=8, 9, 10, 11, 12}
$$
where $S^{AC(2)}_{r}$ is obtained from $S^{AC(1)}_{r}$ by interchanging of
the exponents $l \leftrightarrow m$. Then, from
\beq \label{S81} S_8^{AC(1)} &=& f_8^{(1)}
\sum_{\{n_i\}=0}^\infty \frac
{(y)^{n_2+n_6}(y^{-1})^{n_7}(z)^{n_9}}{n_2!n_6!n_7!n_9!}
\frac{(-1)^{n_7+n_9}(l+\half D|n_2+n_6)}{(1+i+j+\half D|n_9-n_2)} \\[.25cm] 
& \times & \frac{(\s-l|n_7+n_9-n_2)}{(1-i+l|n_2+n_6-n_7-n_9) (1+i+m+\half 
D|n_7-n_6-n_2)} \nonumber ,\eeq 
where 
\beq f_8^{(1)} &=& (-\pi)^D(t^2)^{\s}\;(y^{-1})^{i+m+\half D}\;(z)^{i+j+\half
D}(-i|-j-\half D)(-j|j+l+\half D) \nonumber\\[.25cm] 
& \times & (-l|\s)(-m|-i-\half D)(\s+\half D|i-l-\s-\half D)\nonumber,
\eeq 
we have the general solution
\be S_8^{AC} = S_8^{AC(1)} + S_8^{AC(2)} ,\ee
where the second term is obtained from the first by interchanging
$l\leftrightarrow m$. 

In this solution, it is important to note that there is one pole
(in the special case when the exponents of the propagators are minus
one) of the form $\G(i-l)$. To regularize it one must introduce a small
correction to one of the exponents (a suitable one, of course ), that
is, to take $i=-1-\delta$ and then expand all the factors and the power
series around $\delta=0$. Then, this ``apparent singularity''
cancels out\cite{letb,stand}.

Another result is given by,

\beq \label{S91} S_9^{AC(1)} &=& f_9^{(1)}
\sum_{\{n_i\}=0}^\infty \frac
{(z)^{n_1}(x)^{n_2}(y)^{n_6}(y^{-1})^{n_7}}{n_1!n_2!n_6!n_7!}
\frac{(-1)^{n_1+n_2+n_6+n_7}\;}{(1+i+m+\half
D|n_7-n_2-n_6)} \nonumber\\[.25cm]
& \times & \frac{(l+\half D|n_2+n_6)(m+\half D|n_1+n_7)}{(1-i-j-\half
D|n_1+n_2) (1+j+l+\half D|n_6-n_1-n_7)} ,\eeq  
where 
\beq f_9^{(1)} &=& \pi^D(k^2)^{i+m+D/2}(t^2)^{j+l+D/2}
(-i|-m-\half D)(-j|-l-\half D) \nonumber\\[.25cm]  
& \times &(-l|l+m+\half D)(-m|l+m+\half D)(\s+\half D|-l-m-D) \nonumber,
\eeq 
so that we get,
\be S_9^{AC} = S_9^{AC(1)} + S_9^{AC(2)} .\ee

We note that in this result appears three gamma functions that diverge in four
dimensions. {\it What is the nature of this extra singularity?} It can be a pinch singularity\cite{fairlie}.

Next we have
\beq
\label{S101} S_{10}^{AC(1)} & = & f_{10}^{(1)}\sum_{\{n_i\}=0}^{\infty}\frac
{(z)^{n_2}(z^{-1})^{n_5}(x^{-1})^{n_7}(y^{-1})^{n_8}}{n_2!n_5!n_7!n_8!}\frac
{(-1)^{n_8}(l+\half D|n_2+n_8)}{(1-j+\s|n_7+n_8)}\nonumber\\[.25cm]
& \times & \frac {(m+\half D|n_5+n_7)(i+j+m+D|n_5+n_7-n_2)}{(1+j+m+\half D|n_5-n_2-n_8)},
\eeq
where
\beq 
f_{10}^{(1)} &=& \pi^D (k^2)^{\s} y^j\;(p^2)^{-m-\half D}\;(t^2)^{l+\half
D}(-i|i+m+\half D)(-j|-m-\half D)\nonumber\\[.25cm]
& \times & (-l|2l+\half D)(-m|-i-l-D)(\s+\half D|-l-\half D) \nonumber ,
\eeq
yielding the solution
\be 
S_{10}^{AC} = S_{10}^{AC(1)} + S_{10}^{AC(2)}. \ee


Next we have
\beq 
S_{11}^{AC(1)} &=& f_{11}^{(1)}
\sum_{\{n_i\}=0}^\infty \frac{(z)^{n_2+n_8}(x^{-1})^{n_3}
(z^{-1})^{n_5}(-1)^{n_2+n_3}}{ n_2!n_3!n_5!n_8!
(1+i+l+\half D|n_8-n_3-n_5)}\\[.25cm]  
& \times & \frac{(l+\half D|n_2+n_8)(-j+\s|n_8-n_3)}
{(1-j+l|n_2+n_8-n_3-n_5)(1+j+m+\half D|n_5-n_2-n_8)},\nonumber
\eeq
where
\beq f_{11}^{(1)} &=& \pi^D(p^2)^{i+l+\half D}(t^2)^{j+m+\half
D}(-i|-l-\half D) (-j|\s)\nonumber\\[.25cm]
& \times & (-l|j)(-m|-j-\half D)(\s+\half D|l-\s) \nonumber ,\eeq
which yields the general solution
\be S_{11}^{AC} = S_{11}^{AC(1)} + S_{11}^{AC(2)} .\ee


Finally, we have
\beq S_{12}^{AC(1)} &=& f_{12}^{(1)}
\sum_{\{n_i\}=0}^\infty \frac
{(z)^{n_2}(y^{-1})^{n_3}(z^{-1})^{n_5+n_7}}{n_2!n_3!n_5!n_7!} \nonumber
\\[.25cm]
& \times & \frac{(-1)^{n_3+n_7}(m+\half D|n_5+n_7)}{(1+i+m+\half
D|n_7-n_2-n_3)}\nonumber\\[.25cm]
& \times & \frac{(-l+\s|n_5+n_7-n_2)(-j+\s|n_7-n_3)}{(1+m+\s+\half D|n_5+n_7-n_2-n_3)},
\eeq
where
\beq f_{12}^{(1)} &=& (-\pi)^D(t^2)^{\s}z^{-m-\half D}\;(-i|i+m+\half D)(-j|\s)\nonumber\\[.25cm]
& \times &(-l|\s)(-m|-i-\half D)(\s+\half D|-m-2\s-D) \nonumber ,
\eeq
yielding the general solution
\be 
S_{12}^{AC} = S_{12}^{AC(1)} + S_{12}^{AC(2)} .
\ee 

\subsection{Region of Convergence.}

The various power series we obtained with negative dimensional integration
approach to solve the Feynman integral relative to the chosen two-loop vertex
diagram are very similar to hypergeometric series. Of course hypergeometric
functions of more than two variables are known and they are called
Lauricella's functions\cite{appel}. But in \cite{appel} Appel {\em et
al} studied four Lauricella's functions named
$F_A,\;F_B,\;F_C,\;\mbox{and}\;F_D$ --- even though they mention that
there are other fourteen. So, we do not know what are the regions of
convergence of them nor even how they are called. However, knowing the
region of convergence is not an essential thing here, because our
external legs are off-shell anyway. This question becomes meaningful in
the case where one of the legs are put on-shell, since only in this
regime we can atribute a value to the external momenta. 

\section{On-Shell Limit.}

Of course, particular cases of on-shell external legs must be contained in the
set of off-shell solutions. To check on this, let us take two legs on-shell,
namely, let $k^2=t^2=0$. Not all off-shell solutions $S^{AC}$ are suitable for
taking this particular limit, because some of them either vanish or become
divergent. It is easy to see that such a suitable solution is
given by eq.(\ref{S2}), because in this limit only the first term in
the $F_4$ series is non-zero while all the others vanish, leaving us with
\beq S_2^{AC}(k^2,t^2=0) &=& \pi^D(p^2)^\s (-i|\s)(-j|\s)(-l|-m-\half
D)(-m|l+m+\half D)\nonumber\\[.25cm]
& \times &(l+m+D|-l-\half D)(\s+\half D|-2\s-\half D).
\eeq 

This result is valid for arbitrary $D$ and (negative) exponents of
propagators. In order to confront this result with known one we still
need to go further in specializing to the case where $i=j=l=m=-1$ to
get 
\beq \label{special} S_2^{AC} &=& \pi^D(p^2)^{D-4}\frac
{\G^2(D-3)\G^2(\half D-1)\G(2-\half D) \G(4-D)}
{\G(D-2)\G(\frac{3}{2}D-4)}. \eeq 

This is the very result obtained by Hathrell\cite{hath} using standard
procedures for calculating Feynman diagrams in positive $D$. Of course,
a more straightforward way of getting this result using \ndim{} is to
put the corresponding legs on-shell from the very beginning, and what
we get then is twelve systems to solve with non-trivial solutions,
exactly the number we have for the solution in question: a twelvefold
degeneracy giving the same correct result\cite{lab}. 

We can consider also other special case, namely, the one where
$p^2=k^2=0$. This one is interesting because it contributes to other
two-loop three-point diagram\cite{kramer} if we apply the integration
by parts technique\cite{russo}. The general result, for arbitrary
(negative) exponents of propagators and (positive) dimension can be
read from the solution $S_3$, eq.(\ref{S3}),  

\beq S_3^{AC}(p^2,k^2=0) &=& \pi^D(t^2)^{\s}(-j|\s)(-l|l+m+\half
D)(-m|l+m+\half D)\\[.25cm]  
& \times &(l+m+D|i+j-l-m-\half D)(\s+\half D|-2\s-\half D)\nonumber,
\eeq
taking the same particular case of Kramer\cite{kramer} {\it et al}, i.e.,
$l=m=-2$ and $i=j=-1$, we obtain, 
\beq S_3^{AC} &=& \pi^D(t^2)^{D-6} \frac{\G^3(\half
D-2)\G(D-5)\G(6-D)}{\G(D-4)} ,\eeq 
which is the well-known result in euclidean space.

Two other simpler special cases can be read from this graph, namely, when $p=0,\; k=t$ (see fig.2) and $k=0,\; p=t$
(see fig. 3). The
solution
$S_3$ gives us the first one,

\be S_3^{AC} (p=0,k=t) = f_3 (-i-j|i)(i+l+m+D|-i) ,\ee
note that in the $n_1$ and $n_2$ series only the first term contributes and the $n_3$ one reduces to a gaussian
hypergeometric function with unit argument.

The second case (see fig.3), $k=0,\;p=t$, can be read from $S_2$,

\be S_2^{AC}(k=0,p=-t) = \pi^D (p^2)^\s P_2^{AC} (-i+\s|-\s)(\half D+j-\s|\s) ,\ee
this result can be used to study two self-energy two-loop graphs and agrees with our previous results\cite{flying}.   

Finally, let us check up on a solution that has four series. Let $j=0$, we get
the graph of figure 4. The solution that allows us to consider this limit is
$S_7$,

\be S_7^{AC}(j=0) = \pi^D (k^2)^\s f_7(j=0) ,\ee
observe that there is no sum in the result. This is due to the factor $(-j|n_1+n_2+n_7+n_8)$ which
leads to only one non-vanishing term, i.e., when $n_1=n_2=n_7=n_8=0$ in the
series.

\section{Conclusion.}

We have shown in this paper how we can work out a two-loop vertex
diagram with all external legs off-shell using the \ndim{} technique to
solve the pertinent Feynman integral. Altogether, twenty-one distinct
results are obtained via \ndim{} technique for the considered two-loop
three-point vertex diagram. These are expressed in terms of power
series which can be identified as \hf{}s. The simpler ones are Appel's
$F_4$ \hf{} with two variables, which for the particular cases where
all the exponents of the propagators are set to minus one, can be
transformed into even simpler ones of the gaussian \hf{} type. The
technique provides simultaneously with several \ac{} formulas between
different results, because they arise from the same Feynman integral
(\ref{Indim}). 

\vspace{1cm}

AGMS wishes to thank CNPq (Conselho Nacional de Desenvolvimento
Cient\'{\i}fico e Tecnol\'ogico, Brasil) for financial support.


\vspace{1cm}


\begin{thebibliography}{99} 

\bibitem{zuber}
C.Itzykson, J-B.Zuber, {\it Quantum Field Theory} (McGraw-Hill, 1980).
C.Nash,{\it Relativistic Quantum Fields} (Academic Press, 1978). 

\bibitem{boos}
E.E.Boos, A.I.Davydychev, Theor. Math. Phys. {\bf 89}, 1, (1991)1052.

\bibitem{russo}
K.G.Chetyrkin, A.L.Kataev, F.V.Tkachov, Nucl. Phys. {\bf B174}, (1980) 345.
K.G.Chetyrkin, F.V.Tkachov, Nucl.Phys. {\bf B192} (1981) 159. K.G. Chetyrkin,
hep-th/9610531.

\bibitem{kreimer}
A.Frink, U.Kilian, D.Kreimer, Nucl.Phys.{\bf B488} (1997)426.

\bibitem{thooft}
C.G.Bollini, J.J. Giambiagi, Nuovo Cim. {\bf B12} (1972)20. G.'t Hooft,
M.Veltman, Nucl.Phys.{\bf B44} (1972)189 .

\bibitem{halliday}
I.G.Halliday, R.M.Ricotta, Phys.Lett.{\bf B193}, 2(1987)241.
R.M.Ricotta {\it Topics in Field Theory},(Ph.D. Thesis, Imperial
College, 1987).  

\bibitem{halliday2}
G.V.Dunne, I.G.Halliday, Phys.Lett.{\bf B193}, 2(1987)247.

\bibitem{lab}
A.T.Suzuki, A.G.M.Schmidt, JHEP{\bf 09} (1997) 002. JHEP is an electronic
journal, see for instance its WEB mirrors: {\it http://jhep.sissa.it,
http://jhep.cern.ch }. 

\bibitem{flying}
A.T.Suzuki, A.G.M.Schmidt, Eur.Phys.J. {\bf C5} (1998) 175.

\bibitem{fleischer}
J.Fleischer, V.A.Smirnov, O.V.Tarasov, Z.Phys.{\bf C74} (1997)379.

\bibitem{letb}
N.I.Ussyukina, A.I.Davydychev, Phys.Lett.{\bf B332} (1994)159.

\bibitem{davyd}
A.I.Davydychev, Acta Phys.Polon. {\bf B28} (1997) 841.hep-ph/9610510.

\bibitem{suzuki2}
A.T.Suzuki R.M.Ricotta, {\it XVI Brazilian Meeting on Particles and Fields},
(1995)386, C.O. Escobar (Ed.).

\bibitem{bateman}
A.Erd\'elyi, W.Magnus, F.Oberhettinger and F.G.Tricomi, {\it Higher
Transcendental Functions} (McGraw-Hill, 1953). I.S.Gradshteyn,
I.M.Rhyzik, {\it Table of Integrals, Series and Products} (Academic
Press, 1994).   

\bibitem{rainville}
E.D.Rainville, {\it Special Functions}, (Chelsea Pub.Co., 1960).
P.M.Morse, H.Feshbach, {\it Methods of Theorethical Physics}
(McGraw-Hill, 1953). Y.L.Luke, {\it Special Functions and Their
Approximation} (Vol.I, Academic Press, 1969). 

\bibitem{appel}
P.Appel, J. Kamp\'e de Feriet, {\it Fonctions Hyperg\'eom\'etriques et
Hypersph\'eriques. Polynomes D'Hermite} (Gauthiers-Villars, Paris 1926).

\bibitem{stand}
A.I.Davydychev,  Proc. International Conference "Quarks-92" (World
Scientific, 1993); hep-ph/9307323.  

\bibitem{fairlie}
D.B.Fairlie, P.V.Landshoff, J.Nutall, J.C.Polkinghorne, J.Math.Phys.{\bf 3},
4 (1962) 594. R.J.Eden, P.V.Landshoff, D.I.Olive and J.C.Polkinghorne,
{\it The Analytic S-Matrix} (Cambridge Univ.Press, 1966). I.T.Drummond,
Nuovo Cim.{\bf 29} (1963)720.    

\bibitem{hath}  
S.J.Hathrell, Ann.Phys.{\bf 139} (1982)136.

\bibitem{kramer}
G.Kramer, B.Lampe, J.Math.Phys.{\bf 28} (1987) 945

\end{thebibliography}
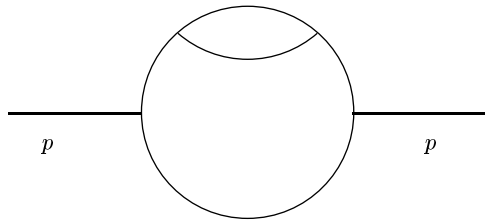
\end{document}